\documentclass[prb,aps,amssymb,twocolumn,showpacs,floatfix]{revtex4}
\usepackage[dvips]{graphicx}
\def\be{\begin{equation}}
\def\ee{\end{equation}}
\def\bea{\begin{eqnarray}}
\def\eea{\end{eqnarray}}
\def\ve{\varepsilon}

\def\bn{{\bf n}}

\def\bj{{\bf j}}
\def\bE{{\bf E}}
\def\bR{{\bf R}}

\def\tq{{\tau_{\rm q}}}
\def\ttr{{\tau_{\rm tr}}}

\def\tin{{\tau_{\rm in}}}
\def\Dd{{D^{\rm (dark)}}}
\def\sd{{\sigma^{\rm (dark)}}}
\def\phid{{\phi^{\rm (dark)}}}
\def\wc{{\omega_c}}
\def\ve{\varepsilon}

\def\w{\omega}

\def\j{{\bf j}}

\begin{document}

\title{Theory of the microwave-induced photocurrent and photovoltage magnetooscillations in a spatially non-uniform 2D electron gas}
\author{I.A.~Dmitriev$^{1,2,*}$}
\author{S.I.~Dorozhkin$^{3}$}
\author{A.D.~Mirlin$^{1,2,\dagger}$}
\affiliation{$^{1}$Institut f\"ur Nanotechnologie, Forschungszentrum
Karlsruhe, 76021 Karlsruhe, Germany\\
$^{2}$Institut f\"ur Theorie der kondensierten Materie,
Universit\"at Karlsruhe, 76128 Karlsruhe, Germany}
\affiliation{$^{3}$ Institute of Solid State Physics, 142432 Chernogolovka, Moscow region, Russia}

\date{\today}

\begin{abstract}
Recent experiment [S.I.~Dorozhkin et al., Phys.\ Rev.\ Lett.\ {\bf 102}, 036602 (2009)] on quantum Hall structures with strongly asymmetric contact configuration discovered microwave-induced photocurrent and photovoltage magnetooscillations in the absence of dc driving. We show that in an irradiated sample the Landau quantization leads to violation of the Einstein relation between the dc conductivity and diffusion coefficient. Then, in the presence of a built-in electric field in a sample, the microwave illumination causes photo-galvanic signals which oscillate as a function of magnetic field with the period
determined by the ratio of the microwave frequency to the cyclotron frequency, as observed in the experiment.
\end{abstract}

\pacs{ 73.50.Pz, 73.43.Qt, 73.50.Fq, 78.67.-n}

\maketitle
\section{Introduction}\label{s1}
\noindent 
Recent developments in the theory of nonequilibrium magnetotransport of a two-dimensional electron gas (2DEG) in high Landau levels are motivated by the discovery of several novel kinds of quantum magnetooscillations induced by microwave radiation,\cite{zudov01}$^{\!-\!}$\cite{highFdu}
 by strong direct current,\cite{yang02,Bykov05,inelasticDC,strongDC,zudov09dc}
 or by phonons.\cite{piro,bykov05,piro08,piro09}
 A particular attention has been attracted by the microwave-induced resistance oscillations (MIRO)\cite{zudov01,ye01} governed by the ratio $\omega/\wc$ of the circular radiation frequency $\w$ and the cyclotron frequency $\wc=|e|B/mc$ (here $B$ is the magnetic field and $m$ the effective
electron mass). Further experiments on MIRO led to spectacular observation of the ``zero resistance states'' (ZRS) in which the dissipative components of both the resistance and conductivity tend to zero.\cite{mani02,zudov03,yang03,dorozhkin03,willett03} These states  were explained in Ref.~\onlinecite{AAM03} as a result of instability leading to formation of domains carrying non-dissipative Hall current.

Initially MIRO were attributed to the  ``{\it displacement}''
mechanism which accounts for spatial displacements of semiclassical electron orbits due to radiation-assisted scattering off disorder.\cite{DSRG03,ryzhii70,VA04,KV08} 
Due to Landau quantization leading to periodic modulation in the density of states (DOS) $\nu(\ve)\simeq\nu(\ve+\wc)$, the preferred direction of such displacements with respect to symmetry-breaking dc field oscillates with $\omega/\wc$. This results in MIRO with the phase and period observed in Refs.~\onlinecite{zudov01,ye01,mani02,zudov03,yang03,dorozhkin03}. 
Later it was realized that the dominant contribution to MIRO in Refs.~\onlinecite{zudov01,ye01,mani02,zudov03,yang03,dorozhkin03} is due to ``{\it inelastic}'' mechanism associated with radiation-induced changes in occupation of electron
states,\cite{DMP03,dorozhkin03,DVAMP0405,DMP07} while the displacement mechanism can be relevant at higher temperatures and only if sufficient amount of short-range impurities is present in the system,\cite{KV08,DKMPV,zudov09} or else, at a very strong dc field\cite{KV08} or microwave power.\cite{DMP07}.

So far, the theoretical research on nonequilibrium magnetooscillations in high Landau levels has been concentrated on the properties of systems which are spatially homogeneous on the macroscopic scale.\cite{DSRG03}$^{\!-\!}$\cite{VAG}
Here we develop more general transport theory applicable also for nonuniform carrier and field distributions. From the experimental side, the present study is motivated by recent experiment\cite{dorozhkin09} which discovered alternating-sign magnetooscillations of photocurrent and photovoltage induced by microwaves in the absence of dc driving. The magnetooscillations with a phase and period similar to MIRO were observed in a 2DEG with a strongly asymmetric contact configuration. The effect was related to the existence of built-in electric fields in a sample in thermodynamic equilibrium,
in particular, in vicinity of doped contacts. As we show below, in an irradiated sample the Landau quantization leads to violation of the Einstein relation between the dc conductivity and diffusion coefficient. Then a finite photocurrent is
driven by a built--in electric field even in the sample at a constant electrochemical potential. In an open circuit, a
photovoltage is produced. Both these photo-galvanic signals
oscillate around zero as a function of magnetic field as observed in the experiment. Another motivation for the present study is the physics of ZRS where the uniform charge and field distributions become electrically unstable, and the knowledge of the transport properties of inhomogeneous system is of central importance for determination of the configuration and dynamics of the current domains.\cite{willett03,Halperin05,Balents05,DorZRS,Halperin09} 

The paper is organized as follows. In next section we formulate an approach applicable for description of the electron kinetics in the presence of non uniformly varying potentials. In Sec.~\ref{s3} we discuss the steady state distributions and current in the absence of the microwave illumination for different experimental setups. In Sec.~\ref{s4} the microwave-induced magnetooscillations in the local transport coefficients are calculated. In Sec.~\ref{s5} we establish the relation between the local transport coefficients and the photocurrent or photovoltage oscillations observed in the experiments. Main findings are summarized in Sec.~\ref{s6}.

\section{Electron kinetics in coordinate and energy space}\label{s2}\noindent 
We consider a 2DEG in a classically strong magnetic field ($\wc\ttr\gg 1$, where $\ttr$ is the transport scattering time),
and in high Landau levels (chemical potential $\mu\gg\wc$). Hereafter we put $\hbar=1$.
Transport of electrons in such system is most conveniently
formulated in terms of migration of the guiding center $\bR(t)$ of the cyclotron
orbits. The dissipative component of the dc current is given by the rate of
changes of $\bR(t)$ due to collisions with impurities 
\be\label{1}
\bj=e n (\partial_t \bR)_{\rm coll}\,,
\ee
where $n$ is the 2D electron density, and $e=-|e|$ is the electron charge.
We describe these collisions using
a generic disorder
model characterized by an arbitrary dependence of the elastic scattering rate 
\be\label{tau_n}
\tau_{\varphi_1-\varphi_2}^{-1}=\sum_{n=-\infty}^\infty \tau_n^{-1} e^{in(\varphi_1-\varphi_2)}, \ \ \ \tau_n=\tau_{-n},
\ee
on the momentum scattering angle $\varphi_2-\varphi_1$. Every scattering event is accompanied by the shift of the guiding center by
$\Delta\bR_{12}=R_c {\bf e}_z\times(\bn_1-\bn_2)$, where $R_c=v_F/\wc$ is the cyclotron radius, $v_F$ the Fermi velocity, and $\bn_k=(\cos\varphi_k,\sin\varphi_k)$ the unit vector in the direction of motion.

Assuming weak one-dimensional spatial variations of $n=n(x)$ and of the
electrostatic potential $\phi=\phi(x)$, we express the migration of the guiding center in terms of the local distribution function $f_{\ve x}$ and
the local density of states (DOS) $\tilde{\nu}_{\ve x}=\nu_0^{-1}\nu[\ve-e\phi(x)]$, 
where $\nu_0=m/2\pi$ is the DOS at $B=0$. In equilibrium, the distribution function $f$ depends only
on the {\it total} energy of electron $\ve$ and is characterized by the position independent electro-chemical potential
$\eta(x)\equiv\mu(x)+e\phi(x)={\rm const}(x)$.\cite{etanote} By contrast, the DOS in high Landau levels is a periodic function of the {\it kinetic} energy $\ve-e\phi(x)$. Generalizing the approach of Refs.~\onlinecite{VA04,KV08,DMP03,DVAMP0405,DMP07,DKMPV} to
the present spatially inhomogeneous case, we obtain
\bea\label{jx}
&&\j_x=2\nu_0 e\int\limits_{-\infty}^x\!dx_1\int\limits_x^\infty\!dx_2(W_{x_1\to x_2}-W_{x_2\to x_1}),\\\nonumber
&& W_{x_1\to x_2}=\langle\ {\cal M}^{\ve_1 \ve_2}_{x_1 x_2}\delta(x_1-x_2+\Delta X_{\varphi_1\varphi_2})
\\\label{W}
&&\times\{\Gamma_{\varphi_1\varphi_2}^{(\rm el)}\delta(\ve_1-\ve_2)+\Gamma_{\varphi_1\varphi_2}^{(\rm ph)}\sum\limits_\pm \delta(\ve_1-\ve_2\pm\w)  \}
\rangle.
\eea
Here $\Delta X_{\varphi_1\varphi_2}={\bf e}_x\cdot\Delta\bR_{12}=R_c(\sin\varphi_1-\sin\varphi_2)$
is the $x$-component of the guiding center shift, factor 2 accounts for the spin degree of freedom, the angular brackets denote averaging over angles $\varphi_{1,2}$ and integrations over $\ve_{1,2}$, and 
\be\label{M}
{\cal M}^{\ve \ve^\prime}_{x x^\prime}=\tilde{\nu}_{\ve x}\tilde{\nu}_{\ve^\prime x^\prime}f_{\ve x}[1-f_{\ve^\prime x^\prime}]\,.
\ee
 The rates of elastic $[\Gamma_{\varphi_1\varphi_2}^{(\rm el)}]$ and photon-assisted 
$[\Gamma_{\varphi_1\varphi_2}^{(\rm ph)}]$ scattering off disorder are given by
\bea\label{el}
\Gamma_{\varphi\varphi^\prime}^{(\rm el)}&=&\frac{1}{\tau_{\varphi-\varphi^\prime}}-\frac{P_{\varphi+\varphi^\prime}}{\tau	_{\varphi-\varphi^\prime}}
\sin^2\frac{\varphi-\varphi^\prime}{2}\,,
\\\label{ass}
\Gamma_{\varphi\varphi^\prime}^{(\rm ph)}&=&\frac{P_{\varphi+\varphi^\prime}}{2\tau_{\varphi-\varphi^\prime}}\sin^2\frac{\varphi-\varphi^\prime}{2}\,.
\eea
The microwave field (screened by the 2D electrons\cite{crossover}) is taken in the form
\be\label{Ew}
\bE_\w(t)=E_\omega \sum_{\pm}{\rm Re}\left[\,s_\pm {\bf e}_\pm e^{i\w t}\right],
\ee
where $2^{1/2}{\bf e}_\pm={\bf e}_x\pm i{\bf e}_y$ and the complex vector $(s_+,s_-)$ of unit length characterizes the polarization. The dimensionless power $P_\theta$ is
\bea\label{Ptheta}
&&P_\theta={\cal P}-2{\rm Re}[{\cal E}_+{\cal E}_-^\star e^{i\theta}],
\\\label{P}
&&{\cal P}=|{\cal E}_+|^2+|{\cal E}_-|^2,
\\\label{Epm}
&&{\cal E}_\pm=s_\pm e v_F E_\w \w^{-1}(\w\pm\wc)^{-1}.
\eea

Apart from the modification of the scattering integral, the microwave illumination leads to a nonequilibrium energy distribution $f_{\ve x}$ of electrons, which is controlled by inelastic relaxation. The corresponding balance equation reads
\be\label{f}
\frac{f_{\ve x}-f^{(T)}_{\ve x}}{\tin}=\langle\tilde{\nu}_{\ve x}^{-1}\Gamma_{\varphi\varphi^\prime}^{(\rm ph)}
\sum_\pm({\cal M}^{\ve\pm\w \ \ve}_{\ x^\prime\ \ x}-
{\cal M}^{\ve \ \ve\pm\w}_{x \ \ x^\prime}
)\rangle_{\varphi\varphi^\prime}\,,
\ee
where $\tin\sim T^{-2}\mu$ is the energy relaxation time due to
electron-electron interaction,\cite{DVAMP0405} ${\cal M}$ is defined in Eq.~(\ref{M}),
$x^\prime=x+\Delta X_{\varphi\varphi^\prime}$, and $f^{(T)}$ is an equilibrium distribution function.
As we will see below, Eqs.~(\ref{jx}) and (\ref{f}) describe both the displacement and the inelastic contributions to photovoltage (or photocurrent) oscillations.\cite{note-mechanisms}

\section{Dark steady state}\label{s3}\noindent 
\subsection{Infinite 2DEG in a constant electric field}\label{ss31}
\begin{figure}[ht]
\centerline{ 
\includegraphics[width=0.95\columnwidth]{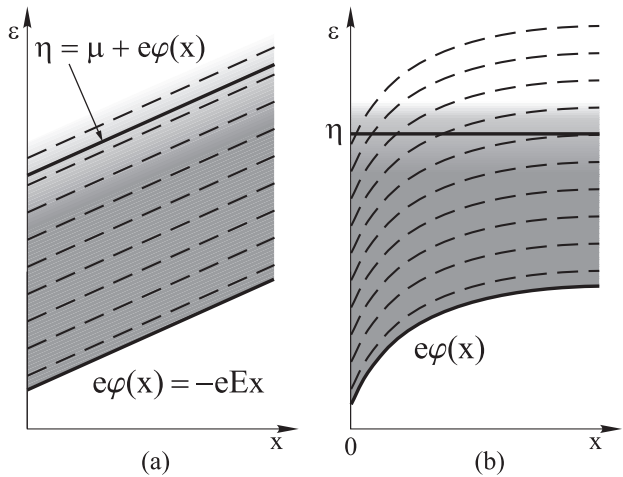}}
\caption{Illustration of the steady state distribution $f_{\ve x}^{(T)}$ (greyscale plot) and position of the Landau levels at $\ve_N(x)=e\phi(x)+(N+1/2)\wc$ (dashed lines) in (a) a non-equilibrium state with a constant electric field $E$ in an infinite 2DEG and of (b) an equilibrium finite 2DEG with a built-in electric field near a contact.}
 \label{fig1}
 \end{figure}

\noindent 
Equations similar to Eqs.~(\ref{jx}), (\ref{f}) were used in Refs.~\onlinecite{DMP03,DVAMP0405,VA04,KV08,DMP07,DKMPV} for analysis of MIRO in a homogeneous case of a constant electric field $E$ in an infinite 2DEG. In this case, the dark (nonequilibrium due to dc current) distribution 
\bea\label{fT}
&&f^{(T)}_{\ve x}=\left(\exp\frac{\ve-\eta(x)}{T}+1\right)^{-1},\\\label{etax}
&&\eta(x)=\mu(x)+e\phi(x)
\eea
is characterized by a coordinate-independent local chemical potential $\mu={\rm const}(x)$. The occupation of
all states having equal {\it kinetic} energy $\ve-e\phi(x)=\ve+eEx$ is the same, see
Fig.~\ref{fig1}a. 
\subsection{Inhomogeneous equilibrium state}\label{ss32}\noindent 
Let us now consider an equilibrium 2DEG in the absence of the microwave field but in the presence of a built-in static electric field. 
In the absence of external voltage applied to the sample, this static field can be created, for instance, by a metallic contact, as in Fig.~\ref{fig1}b. The distribution function
(\ref{fT}) in this inhomogeneous equilibrium case is characterized by a position-independent electrochemical potential 
\be\label{eta}
\eta={\rm const}(x),
\ee
while both electron concentration and electrical potential vary with $x$. 
By contrast to the previous case of a constant electric field, here all states having equal {\it total}
energy $\ve$ are equally occupied, $f^{(T)}_{\ve x}=f^{(T)}_\ve$. 

In the absence of the microwave field, $E_\omega=0$, the inelastic
part $\propto \Gamma_{\varphi_1\varphi_2}^{(\rm ph)}=0$ of the total current (\ref{jx}) is absent while the elastic
part proportional to ${\cal M}^{\ve \
\ve}_{x^\prime x}-{\cal M}^{\ve \ \ve}_{x \ x^\prime}=0$ vanishes as 
$f_{\ve x^\prime}=f_{\ve x}$, see Eq.~(\ref{M}).
We arrived at the trivial result that the current does not't flow when 
the 2DEG is in the equilibrium state characterized by a given 
temperature $T$ and electrochemical potential $\eta$.

\subsection{Electrical current and diffusion; Einstein relation}\label{ss33}\noindent 
Vanishing of electrical current in an equilibrium state of inhomogeneous system can be equivalently formulated as the Einstein relation between the 
linear--response conductivity and diffusion coefficient.
It is instructive to derive this relation by considering weak perturbations of a spatially homogeneous equilibrium system with a fixed concentration of electrons $n={\rm const}(x)$ [and, therefore, $\mu={\rm const}(x)$]. 

According to Eq.~(\ref{jx}), the electric current induced in this system by 
the infinitesimally small electric field $E=-\nabla\phi$ reads
\bea\label{jE}
&&j^{\rm (dark)}|_{\nabla n=0}=-2\sd \nabla\phi\,,\\\label{sd}
&&\sd=-\sigma_{\rm D}\int\!d\ve\tilde{\nu}_\ve^2
\partial_\ve f^{(T)}, 
\eea
Here $\sigma_{\rm D}=e^2\nu_0 R_c^2/2\ttr$ is the classical Drude conductivity per spin orientation in a strong $B$ ($\wc\ttr\gg1$),  the transport relaxation time $\ttr$
is expressed in terms of the moments $\tau_n$, Eq.~(\ref{tau_n}), as $\ttr=(\tau_0^{-1}-\tau_1^{-1})^{-1}$, and the superscript ``(dark)'' refers to the equilibrium state in the absence of  microwaves. 

Now we put $E=-\nabla\phi=0$ and calculate the diffusion current, i.e. the linear response to a small gradient $\nabla n$ of the concentration
\be\label{n}
n(x)=2 	\nu_0\int d\ve \tilde{\nu}(\ve) f_{\ve x}\,.
\ee
The diffusion current 
\be
j^{\rm (dark)}|_{\nabla \phi=0}=-e \Dd \nabla n\,,\label{jn}
\ee
defines the dark diffusion coefficient 
\be\label{Dd}
\Dd=2\nu_0 {\cal{D}}^{\rm (dark)}/ \chi^{\rm (dark)}\,.
\ee
Using $n$ given by Eq.~(\ref{n}), we express the dark compressibility as
\be\label{chi} 
\chi^{\rm (dark)}=\partial n/\partial\mu
=-2 \nu_0\int d\ve \tilde{\nu}(\ve) \partial_\ve f^{(T)}.
\ee 
The quantity ${\cal{D}}^{\rm (dark)}$ has the dimensionality of the diffusion coefficient $D^{\rm (dark)}$ and 
is defined through the relation 
\be\label{jmu}
j^{\rm (dark)}|_{\nabla\phi=0}=-2 e \nu_0 {\cal D}^{\rm (dark)} \nabla \mu(x)
\ee
as the current response to the gradient of the chemical potential $\nabla \mu(x)$.
Calculation using Eq. ~(\ref{jx}) gives
\be\label{calDd}
{\cal{D}}^{\rm (dark)}=-\frac{R_c^2}{2\ttr}\int\!d\ve\tilde{\nu}_\ve^2 \partial_\ve f^{(T)}\,.
\ee

If we now allow for a generic weak perturbation of a homogeneous equilibrium system, the local current takes the form
\be\label{nomw}
j^{\rm (dark)}=-2\sd \nabla\phi(x)-e \Dd
\nabla n(x)\,.
\ee
According to Eqs.~(\ref{jE})-(\ref{calDd}), the diffusion coefficient $\Dd$ and conductivity $2\sd$ are related as
\be
\label{einstein}
2\sd=e^2 \chi^{\rm (dark)} \Dd\,.
\ee
The Einstein relation (\ref{einstein}) ensures the absence of the electron flow in the equilibrium state (\ref{eta}) with 
\be\label{nablaeta}
\nabla\eta=\nabla\mu+e\nabla\phi=(\chi^{\rm (dark)})^{-1}\nabla n+e\nabla\phi=0.
\ee

Related to Eq.~(\ref{einstein}) is a simpler identity 
\be
\label{einstein1}
\sd=e^2 \nu_0 {\cal D}^{\rm (dark)}\,,
\ee
which does not involve the compressibility.

An important consequence of the Einstein relation 
is that 
the current response of any equilibrium system can be represented as
\be\label{jeta}
j^{\rm (dark)}=-2 e \nu_0 {\cal D}^{\rm (dark)} \nabla\eta(x)\,,
\ee
i.e. the current is proportional to the gradient of the electrochemical potential independently of what kind of perturbation causes the current flow.

In what follows we assume an experimentally relevant range of
high temperatures, $2\pi^2 T/\wc\gg 1$, where Shubnikov - de Haas oscillations
are thermally suppressed and transport properties are independent of the
position of the chemical potential with respect to Landau levels.\cite{etanote} In this limit,
Eq.~(\ref{sd}) reduces to
\be
\label{sd2}
\sd=\sigma_{\rm D} \langle\tilde{\nu}_\ve^2\rangle_\ve\,,\qquad 2\pi^2 T/\wc\gg 1.
\ee
Here $\langle\ldots\rangle_\ve$ implies energy averaging over the
$\wc$-periodic DOS oscillations. In high-$T$ limit, the dark compressibility (\ref{chi}) reduces to $2\nu_0$, 
\be\label{chi2}
\chi^{\rm (dark)}=2\nu_0\langle\tilde{\nu}_\ve\rangle_\ve=2\nu_0\,,\qquad 2\pi^2 T/\wc\gg 1.
\ee
A simple linear relation $n(x)=2\nu_0\mu(x)$ makes two definitions of the diffusion current (\ref{jn}) and (\ref{jmu}) identical,
\be\label{DD}
D^{\rm (dark)}\equiv{\cal D}^{\rm (dark)}=\frac{R_c^2}{2\ttr}\langle\tilde{\nu}_\ve^2\rangle_\ve\,,\qquad \frac{2\pi^2 T}{\wc}\gg 1.
\ee

In the presence of microwaves, however, the two definitions are not fully 
equivalent in view of 
the microwave-induced magnetooscillations in the compressibility (MICO),\cite{compress} see discussion in Sec.~\ref{ss43} and in Sec.~\ref{ss51}.

\section{Local conductivity and diffusion coefficient in illuminated 2DEG}\label{s4}\noindent 
\subsection{Nonequilibrium current flow }\label{ss41}\noindent 
We now turn to evaluation of the transport properties in the presence of microwave radiation. 
The key observation is that in the nonequilibrium steady state the Einstein relation (\ref{einstein1}) between the dc conductivity $\sigma$ and diffusion coefficient ${\cal D}$ does
not hold anymore, $\sigma\neq e^2\nu_0{\cal D}$. In other words, the current cannot be represented in the form of Eq.~(\ref{jeta}) with some modified transport coefficient and electrochemical potential. According to our calculation based on Eqs.~(\ref{jx}) and (\ref{f}), the nonequilibrium dc current 
\be\label{genj}
j_x=-2\sigma_\eta\nabla\phi-2e\nu_0 {\cal D}\nabla\eta.
\ee
necessarily contains an extra "anomalous term" $-2\sigma_\eta\nabla\phi$ violating the Einstein law.  In these terms, the total conductivity $\sigma$, which defines the dc current
$j=2\sigma E$ in a homogeneous system (as in the case of MIRO, see Sec.~\ref{ss31}), is given by
\be\label{sigma}
\sigma=\sigma_\eta+e^2\nu_0 {\cal D},
\ee
while the diffusion coefficient $D$ entering the current $j|_{E=0}=-e D\nabla n$ 
at $E=0$, is expressed through the nonequilibrium compressibility $\chi$,\cite{compress}
\be\label{DDD}
D=2\nu_0 \chi^{-1}{\cal D}\,,
\ee
similar to Eq.~(\ref{Dd}). In next two subsections we calculate the anomalous conductivity $\sigma_\eta$ and a photoinduced part of the diffusion coefficient to the minimal order $E_\w^2$.

\subsection{Anomalous component of conductivity}\label{ss42}\noindent 
In this subsection we calculate the anomalous component $\sigma_\eta$ of the conductivity. For that purpose we put $\nabla\eta=0$ and use Eqs.~(\ref{jx}) and (\ref{f}) with the position-independent dark distribution,
\be \label{feta}
f^{(T)}_\ve=[e^{(\ve-\eta)/T}+1]^{-1},\ \ \eta={\rm const}(x).
\ee
Similar to the dark case, Sec.~{\ref{ss32}}, the microwave correction (\ref{el}) to the elastic scattering rate gives no
contribution to the current (\ref{jx}) due to cancellation ${\cal M}^{\ve \
\ve}_{x^\prime x}-{\cal M}^{\ve \ \ve}_{x \ x^\prime}=0$. 
Therefore, the current
$\propto E_\w^2$ can be either due to (i) the microwave-assisted scattering off
disorder (represented by the second term $\propto\Gamma^{(\rm ph)}$ in
Eq.~(\ref{jx}) with unperturbed $f=f^{(T)}_\ve$, displacement mechanism) or due
to (ii) the position dependence of the microwave-induced nonequilibrium distribution (\ref{f}) [modifying the
elastic term in the total current (\ref{jx}), inelastic mechanism]. Correspondingly, the anomalous conductivity $\sigma_\eta$ is
a sum of the displacement and inelastic contributions
\be
\label{sigmaeta}
\sigma_\eta=\sigma^{(\rm dis)}_\eta+\sigma^{(\rm in)}_\eta
\ee

{\it Displacement contribution} $\sigma^{(\rm dis)}_\eta$. Using the position-independent distribution (\ref{feta}), the second term of Eq.~(\ref{jx}) can be represented as
\bea\nonumber
&&j_\eta^{(\rm dis)}=2e\nu_0\sum_\pm\int\!d\ve (f^{(T)}_{\ve}-f^{(T)}_{\ve\pm\w})
\langle \Theta(\Delta)\Gamma_{\varphi\varphi^\prime}^{(\rm ph)}\\
\label{dis}&&\times\!\!
\int\limits_{x-\Delta}^x\!\!\!dx^\prime\,\tilde{\nu}[\ve-e\phi(x^\prime)]\tilde{\nu}[\ve\pm\w -e\phi(x^\prime+\Delta)]\rangle_{\varphi\varphi^\prime}.
\eea
where the Heaviside function $\Theta(\Delta)$ imposes the condition $\Delta\equiv\Delta X_{\varphi\varphi^\prime}>0$, see Eq.~(\ref{jx}). The electric field enters this expression only through the position dependence of the local DOS $\tilde\nu_{\ve x}=\tilde\nu[\ve-e\phi(x)]$. In the absence of the local electric field $E=-\nabla\phi(x)$, two terms of Eq.~(\ref{dis}) corresponding to $\ve\pm\w$ exactly cancel each other. The terms linear in $E$ produce the displacement contribution to the current $j_\eta^{(\rm dis)}=\sigma_\eta^{(\rm dis)}E$ under condition (\ref{eta}):
\bea\label{sigmadis}
&&\sigma_\eta^{(\rm dis)}=\sigma_{\rm D}\frac{\ttr}{4\tau_\star}({\cal P}-{\rm Re}[{\cal
E}_+{\cal E}_-^\star]){\cal R}_1(\w),
\\\label{R1}&&{\cal R}_1(\w)=\w\partial_\w\langle\tilde{\nu}_\ve\tilde{\nu}_{\ve+\w}\rangle_\ve.
\eea
Here we used $2\pi^2 T/\wc\gg 1$, and 
\be\label{tau*}
\tau_\star^{-1}=3\tau_0^{-1}-4\tau_1^{-1}+\tau_2^{-1}
\ee
 in terms of Eq.~(\ref{tau_n}). Function ${\cal R}_1(\w)$ oscillating with the ratio $\w/\wc$ is specified 
in Sec.~\ref{ss44} for two limits of strongly overlapping and well separated Landau levels, together with similar oscillatory functions entering Eqs.~(\ref{R2}) and (\ref{R3}).

{\it Inelastic contribution} $\sigma^{(\rm in)}_\eta$. The inelastic contribution is due to microwave-induced changes $\delta f$ in the distribution function. To the leading order $E_\w^2$ and in the limit $\nabla\phi(x)\to0$, Eqs.~(\ref{f}) and  (\ref{eta}) give
\be\label{f1}
\delta\! f_{\ve x}={\cal P}\frac{\tin}{4\ttr}\sum\limits_\pm(f^{(T)}_\ve-f^{(T)}_{\ve\pm\w})\tilde{\nu}[\ve\pm\w-e\phi(x)].
\ee
In contrast to the spatially independent $f^{(T)}_\ve$, the microwave-induced
part $\delta f$ of the electronic distribution oscillates in coordinate space at
a fixed total energy $\ve$ due to spatial oscillations of DOS $\tilde{\nu}_{\ve
x}$ in Landau levels tilted by the electric field. As a result, the elastic
contribution to the current (\ref{jx}) does not vanish, ${\cal M}^{\ve \
\ve}_{x^\prime x}-{\cal M}^{\ve \ \ve}_{x \ x^\prime}\neq 0$. 
Substitution of Eq.~(\ref{f1}) for $f_{\ve x}$ in the elastic
term of  Eq.~(\ref{jx}) with
$\Gamma_{\varphi\varphi^\prime}^{(\rm el)}=\tau_{\varphi-\varphi^\prime}^{-1}$
produces inelastic contribution to the current,
\be\label{jin}
j^{(\rm in)}_\eta\!=\!2e\nu_0\int\!d\ve\left\langle\!\Delta\Theta(\Delta)\!\!\int\limits_
{ x-\Delta}^x\!\!\!dx^\prime\,\frac{\tilde{\nu}_{\ve
x^\prime}^2\nabla_{x^\prime}\delta\!f_{\ve
x^\prime}}{\tau_{\varphi-\varphi^\prime}}\!\right\rangle_{\varphi\varphi^\prime},
\ee
where $\Delta\equiv\Delta X_{\varphi\varphi^\prime}>0$ as above in Eq.~(\ref{dis}). Assuming $2\pi^2 T/\wc\gg 1$ and keeping the linear term   $\propto\nabla\phi(x)$, we obtain $j_\eta^{(\rm in)}=\sigma_\eta^{(\rm in)}E$ with
\bea\label{sigmain}
&&\sigma_\eta^{(\rm in)}=\sigma_{\rm D}\frac{\tin}{4\ttr}{\cal P}{\cal R}_2(\w),
\\\label{R2}&&{\cal R}_2(\w)=\w\partial_\w\langle\tilde{\nu}^2_\ve(\tilde{\nu}_{\ve+\w}+\tilde{\nu}_{\ve-\w})\rangle_\ve.
\eea

It is worth mentioning that both $\sigma_\eta^{(\rm dis)}$ and $\sigma_\eta^{(\rm in)}$ originate from the spatial dependence of the DOS which requires both the Landau quantization and the presence of electric field. In the absence of Landau quantization, $\tilde{\nu}=1$,  functions ${\cal R}_1(\w)$ and ${\cal R}_2(\w)$ entering Eqs.~(\ref{R1}) and (\ref{R2}) vanish (see Sec.~\ref{ss44} ). Therefore, within our model the Einstein relation of the dc conductivity and diffusion coefficient is restored in the classical\cite{classical} limit $\wc\tau_0\to 0$: $\sigma_\eta=\sigma_\eta^{(\rm dis)}+\sigma_\eta^{(\rm in)}=0$, see Eqs.~(\ref{genj}), (\ref{sigmaeta}), (\ref{sigmadis}), (\ref{sigmain}), and (\ref{RROvLLs}).

\subsection{Microwave-induced oscillations of the diffusion coefficient}\label{ss43}\noindent 
Now we assume $\nabla\phi=0$ and $\nabla\mu=\nabla\eta\neq 0$, and calculate the microwave-induced correction $D_{\rm ph}=D-\Dd$ to the diffusion coefficient, see Eqs.~(\ref{genj}) and (\ref{DDD}). In contrast to the previous subsection, now the DOS is position independent, while the dark distribution varies in space.  Using the linear approximation $f^{(T)}_{\ve x+\delta x}=f^{(T)}_{\ve x}-\delta x\,\nabla\mu\,\partial_\ve f^{(T)}_{\ve x}$ in Eq.~(\ref{jx}), we obtain $j_x=-2e\nu_0 {\cal D}\nabla\mu$ with
\bea\nonumber
{\cal D}&=&\int\!d\ve [-\partial_\ve f^{(T)}_{\ve x}]\langle \Theta(\Delta X_{\varphi\varphi^\prime})\Delta X_{\varphi\varphi^\prime}^2\\
\label{D1}
&\times&[\tilde{\nu}_\ve^2\Gamma_{\varphi\varphi^\prime}^{(\rm el)}+\tilde{\nu}_\ve(\tilde{\nu}_{\ve-\w}+\tilde{\nu}_{\ve+\w})
\Gamma_{\varphi\varphi^\prime}^{(\rm ph)}]\rangle_{\varphi\varphi^\prime}\,.
\eea
Performing the angular and thermal averaging for $2\pi^2 T/\wc\gg 1$, we get, similar to Eqs.~(\ref{sigmadis})--(\ref{tau*}),
\bea\label{Ddis}
&&{\cal D}_{\rm ph}=\frac{R_c^2}{8\tau_\star}({\cal P}-{\rm Re}[{\cal
E}_+{\cal E}_-^\star]){\cal R}_3(\w),
\\\label{R3}&&{\cal R}_3(\w)=\langle\tilde{\nu}_\ve^2-\tilde{\nu}_\ve\tilde{\nu}_{\ve+\w}\rangle_\ve\,,
\eea
where ${\cal D}_{\rm ph}={\cal D}-{\cal D}^{\rm (dark)}$ and ${\cal D}^{\rm (dark)}$ is given by Eq.~(\ref{DD}).

In Eqs.~(\ref{D1}) and (\ref{Ddis}), the microwave-induced changes in the distribution function (\ref{f1}) were not taken into account. The reason is that in the limit $2\pi^2 T/\wc\gg 1$ the corresponding contribution to the diffusion coefficient is exponentially suppressed. The inelastic contribution ${\cal D}^{(\rm in)}_{\rm ph}$, obtained from Eqs.~(\ref{f1}) and (\ref{jin}) using $\nabla\phi=0$ and $\nabla\mu\to 0$, reads
\be\label{Din}
{\cal D}^{(\rm in)}_{\rm ph}=\frac{{\cal P}\tin R_c^2}{8\ttr^2}\sum\limits_\pm\int\!d\ve\, \tilde{\nu}_\ve^2\,\tilde{\nu}_{\ve\pm\w}\,
\partial_\ve [f^{(T)}_{\ve x}-f^{(T)}_{\ve\pm\w x}].
\ee
In the limit $2\pi^2 T/\wc\gg 1$, this expression vanishes similar to the Shubnikov-de Haas oscillations. Therefore, the microwave-induced oscillations in the distribution function (\ref{f1}) produce the contribution (\ref{sigmain}) to the anomalous conductivity $\sigma_\eta$ only, while the displacement mechanism provide similar oscillations both in $\sigma_\eta$, Eq.~(\ref{sigmadis}), and in ${\cal D}$, Eq.~(\ref{Ddis}).  

While the dark quantities ${\cal D}^{\rm (dark)}$ and $D^{\rm (dark)}$ are identical at  $2\pi^2 T/\wc\gg 1$, see Eq.~(\ref{DD}),
in the presence of microwaves they are not equivalent in view of 
the microwave-induced compressibility oscillations (MICO).\cite{compress} 
MICO do not enter the quantity ${\cal D}$  since it is defined through $\nabla\mu$, but modify the diffusion coefficient $D$ defined through $\nabla n=\chi\nabla\mu$, which gives
\be\label{DDD1}
D_{\rm ph}\equiv D-D^{\rm (dark)}=2\nu_0 \chi^{-1}{\cal D}_{\rm ph}\,.
\ee
However, since we are interested in linear-in-$E_\w^2$ corrections to $D^{\rm (dark)}$
and ${\cal D}^{\rm (dark)}$ in the present work, we can approximate the compressibility by its dark value $\chi^{\rm (dark)}=2\nu_0$ (thus neglecting MICO that lead to terms $\propto E_\w^4$ in $D_{\rm ph}$). Moreover, even at high orders in $E_\w$, the compressibility can be approximated as $\chi=\partial n/\partial\mu=2\nu_0$ assuming spatial variations of $n(x)$ are smooth on a scale of the inelastic length. At shorter length scales, MICO can be strong (of order $\nu_0$). This situation arises, in particular, in the regime of ZRS,\cite{compress} see Sec.~\ref{ss51}.

So far we considered the two cases $\nabla\eta=0$ and $\nabla\phi=0$ which give,
correspondingly, the anomalous conductivity $\sigma_\eta$ and the photoinduced
correction to the diffusion coefficient $D=\Dd+D_{\rm ph}$. The sum $\sigma_{\rm
ph}=\sigma_\eta+e^2\nu_0 {\cal D}_{\rm ph}$, given by Eqs.~(\ref{sigmaeta}),
(\ref{sigmadis}), (\ref{sigmain}), and (\ref{Ddis}), reproduces the
results\cite{VA04,KV08,DMP03,DVAMP0405,DMP07,DKMPV}
obtained earlier for the homogeneous case of the MIRO,\cite{zudov01,ye01,mani02,zudov03,yang03,dorozhkin03} which corresponds to $\nabla\mu=0$ and to the constant $-\nabla\phi=-\nabla\eta/e=E$,  see also Sec.~\ref{ss53}.

\subsection{Form of the oscillations for overlapping and separated Landau levels}\label{ss44}\noindent 
The form and the phase of the magnetooscillations in the anomalous conductivity $\sigma_\eta$ and in the diffusion coefficient $D$, Eqs.~(\ref{R1}), (\ref{R2}), and (\ref{R3}), as well as the quantum correction to the dark conductivity, Eqs.~(\ref{sd2}), are expressed through the certain energy averages ${\cal R}_n(\w)$ over the period $\wc$ of the DOS oscillations, 
\bea\nonumber
&&{\cal R}_0=\langle\tilde{\nu}_\ve^2\rangle_\ve,\\\nonumber
&&{\cal R}_1(\w)=\w\partial_\w\langle\tilde{\nu}_\ve\tilde{\nu}_{\ve+\w}\rangle_\ve,\\\nonumber
&&{\cal R}_2(\w)=\w\partial_\w\langle\tilde{\nu}_\ve^2(\tilde{\nu}_{\ve+\w}+\tilde{\nu}_{\ve+\w})\rangle_\ve,\\
&&{\cal R}_3(\w)=\langle\tilde{\nu}_\ve^2-\tilde{\nu}_\ve\tilde{\nu}_{\ve+\w}\rangle_\ve.\label{RR}
\eea
Here we specify these functions in two limits of strongly overlapping and of well-separated Landau levels (LLs) within the self-consistent
Born approximation (SCBA).
At high LLs, $\ve_F\gg \omega,\omega_c$, disorder can be treated within the SCBA provided the
disorder correlation length satisfies $d\ll l_B$ and $d\ll v_F\tau_{\rm q}$, where $l_B=(c/eB)^{1/2}$ is the magnetic length and $\tau_{\rm q}$ the quantum relaxation time [$\tau_{\rm q}\equiv\tau_0$ in terms of the moments $\tau_n$, Eq.~(\ref{tau_n})].\cite{SCBA,DMP03,VA04,KV08}

In moderate magnetic field, $\delta=\exp(-\pi/\omega_c \tau_{\rm q})\ll 1$, LLs strongly overlap and the DOS is only weakly modulated by the magnetic field,   $\tilde{\nu}(\ve)=1-2\delta\cos(2\pi\ve/\omega_c)$. In this limit
\bea\nonumber
&&{\cal R}_0=1+2\delta^2,\\\nonumber
&&{\cal R}_1=-2\delta^2\frac{2\pi\w}{\wc}\sin\frac{2\pi\w}{\wc},\\\nonumber
&&{\cal R}_2=-8\delta^2\frac{2\pi\w}{\wc}\sin\frac{2\pi\w}{\wc},\\
&&{\cal R}_3=4\delta^2\sin^2\frac{\pi\w}{\wc}.\label{RROvLLs}
\eea

In the limit of separated
LLs, $\wc\tau_{\rm q}\gg 1$, the DOS is a sequence of semicircles of
width $2\Gamma=2(2\wc/\pi\tau_{\rm q})^{1/2}\ll\wc$, i.e.,
$\tilde{\nu}(\ve)=\tau_{\rm q}{\rm Re}\sqrt{\Gamma^2-(\delta\ve)^2}$,
where $\delta\ve$ is the detuning from the center of the nearest
LL. In this limit, calculation yields
\bea\label{R0sep}
&&{\cal R}_0=16\wc/3\pi^2\Gamma,\\\label{R1sep}
&&{\cal R}_1={\cal R}_0{\w\over\Gamma}\sum\limits_n{\rm sgn}(\Omega_n){\cal
H}_2(|\Omega_n|)\,,\\\label{R2sep}
&&{\cal R}_2=-{\cal R}_0{4\w\wc\over\Gamma^2}
\sum\limits_n{\rm sgn}(\Omega_n) \Phi_2(|\Omega_n|)\,\\
&&{\cal R}_3={\cal R}_0\left[1-\sum\limits_n{\cal H}_1(|\Omega_n|)\right].\label{R3sep}
\eea
The parameterless functions of $\Omega_n=(\w-n\wc)/\Gamma$ are nonzero at
$0\!<\!|\Omega_n|\!<\!2$, where they are expressed as
\bea\label{H1}
{\cal H}_1(x)\!&=&\!(2+x)\,[\,(4+x^2)E(X)-4x K(X)\,]/8\,,\qquad
\\\label{H2}
{\cal H}_2(x)\!&=&\!3x\,[\,(2+x)E(X)-4K(X)\,]/8\,,
\\\label{Phi2}
4\pi\,\Phi_2(x)\!&=&\! 3x\,{\rm arccos}(x-1)-x(1+x)\sqrt{x(2-x)}\,.
\eea
Here $X\equiv(2-x)^2/(2+x)^2$ and the functions $E$ and $K$ are the complete elliptic
integrals of the first and second kind, respectively. Graphical representation of the functions (\ref{H1})-(\ref{Phi2}) can be found in Ref.~\onlinecite{FMIRO}. 

In the crossover magnetic field, $\wc\tau_{\rm q}\sim 1$, 
functions (\ref{RR}) obtained using analytical expressions for the DOS become very cumbersome. 
In this crossover region, the form of the oscillations can be obtained using numerical 
solution of the SCBA equations.\cite{crossover} In particular, such numerical 
solution is used in the calculation illustrated in Fig.~\ref{fig2}b below.

\section{Photocurrent and photovoltage oscillations}\label{s5}\noindent 
In this section, we use the obtained local transport coefficients 
for  calculation of the electrical current in different experimental situations. The effects related to microwave-induced modifications 
of the spatial
distribution of carriers and fields are considered in  Sec.~\ref{ss51}. The current-voltage characteristics  (CVC) of an infinite 2DEG stripe between two metallic contacts are obtained in Sec.~\ref{ss52}. Using these CVC, in Sec.~\ref{ss53} we calculate the photocurrent and photovoltage  and compare our findings with the experiment of Ref.~\onlinecite{dorozhkin09}. In Sec.~\ref{ss54}, nonlinear effects in the photovoltage (with respect to the microwave power) are discussed, which were observed experimentally\cite{dorozhkin09} and are also well reproduced by the theory.

\subsection{Photo-induced changes in the field and charge distribution}\label{ss51}\noindent 
In the presence of nonuniform carrier and field distributions, as in
Fig.~\ref{fig1}b, the local transport coefficients $\sigma_\eta$
and ${\cal D}$ entering the local current (\ref{genj}) do not completely determine
the transport. The full theory should include a self-consistent solution of
the Poisson and continuity equations for a given experimental setup. Indeed, the
photoinduced current density, given by Eq.~(\ref{genj}) with the dark
profile of the electrostatic potential $\phid$ and with $\nabla\eta=0$, is
$j_x=-2\sigma_\eta\nabla\phid$. In general, such $j_x$ does not satisfy the continuity
equation $\nabla j_x=0$ in view of a nonlinear spatial variation of $\phid$, as, for instance, in
Fig.~\ref{fig1}b. Therefore, the photoinduced variation of the electron density 
$\delta n(x)=n-n_{\rm dark}$ (we assume $|\delta n(x)|\ll n$) and of the electrostatic potential
$\delta\phi(x)=\phi-\phid$ should be taken into account. The latter are
related to each other  by the inverse capacitance matrix $\widehat{W}$ as
\be
\delta\phi_x=e\int\!dx\,W_{xx^\prime}\delta n_{x^\prime}.
\ee
 Using the relation
\be
\delta\eta=e\delta\phi +\chi^{-1}\delta n
\ee
valid at $2\pi^2 T/\wc\gg 1$, 
we represent the Poisson equation in the form
\be\label{Poisson}
\delta\eta=[1+(e^2\chi\widehat{W})^{-1}]e\delta\phi.
\ee 
Using Eqs.~(\ref{genj}) and (\ref{Poisson}) for a fixed current density 
$j_x=j={\rm const}(x)$, one arrives to a formal solution for the local variation
$\delta\eta(x)$ of the electrochemical potential,
\bea\nonumber
\delta\eta&=&-e\left\{2\sigma_\eta\nabla[1+(e^2\chi\widehat{W})^{-1}]^{-1}+ 2e^2\nu_0 {\cal D} \nabla \right\}^{-1}\\\label{etaP}
 &\times&(j+2\sigma_\eta\nabla\phid).
\eea

Solution of the above non-local equation is required if the amplitude of
oscillations in $\sigma_\eta$ becomes of order $\sd=e^2\nu_0
{\cal D}^{\rm (dark)}$ [otherwise one can neglect the photoinduced changes $\delta n(x)$ in view
of the smallness of $\sigma_\eta\propto E^2_\w$]. In conventional
magnetoresistivity experiments this corresponds to the regime where the
zero-resistance states are
formed\cite{willett03,Halperin05,Balents05,DorZRS,Halperin09,compress} 
[both theory and experiments show that the ZRS appear still in
the linear regime in the microwave power where Eqs.~(\ref{sigmaeta}),
(\ref{sigmadis}), (\ref{sigmain}), and (\ref{Ddis}) still apply].
According to the theory of Ref.~\onlinecite{AAM03}, the ZRS is a manifestation of a
spontaneous symmetry breaking of a homogeneous state with negative resistivity
leading to the formation of the current domains. In this picture, the residual
resistivity in the ZRS, which is observed in part of experiments, is due to the
electron transport across the domain walls and near the boundary of the 2DEG. 
Inside the domains, the transport is dissipationless. 

The boundaries of the domains are characterized by strongly nonuniform carrier
and field distributions. Therefore, the results of the present work, in particular,
the violation of the Einstein relation and the appearance of the anomalous
component of conductivity $\sigma_\eta$, should play an important role for 
development of microscopic theory of transport in the ZRS regime.

\subsection{Boundary conditions and current-voltage characteristics}\label{ss52}\noindent
We now consider the photocurrent and photovoltage oscillations in a 2DEG with metallic contacts.
As we show below, specific boundary conditions (\ref{bc}) at the interface with 
metallic contacts make the details of the potential and carrier 
distributions inside the sample irrelevant [thus, one need not solve a complicated electrostatical problem (\ref{etaP})]. 
More precisely, as long as simple 1D or Corbino geometry is considered (see Fig.~\ref{fig3}) and the built-in electric field is not too strong , the
current and voltage between the contacts are fully determined by the difference of the work
functions of the contacts and by the local transport coefficients
$\sigma_\eta$ and ${\cal D}$, see Eq.~(\ref{CVC}) below.

Indeed, using the fact that in the
linear approximation with respect to the dc field not only $j_x$ but also
$\sigma_\eta$ and ${\cal D}$ are position independent, and integrating both parts of
Eq.~(\ref{genj}) along a contour connecting two
contacts at $x=0$ and $x=L$, we obtain the relation 
\be\label{curgen}
j L=2\sigma_\eta[\phi(0)-\phi(L)]+2e\nu_0 {\cal D}[\eta(0)-\eta(L)].
\ee
It is natural to assume that microwave radiation does not change the
electron concentration on the metallic side of the interfaces due to a huge
density of states there. Since both the electrochemical and electrostatic
potentials are continuous at the interface, this fixes the chemical potential in
the 2DEG near the interfaces. Introducing the voltage
$V=[\eta(0)-\eta(L)]/e$ and the difference of the
work functions of the two contacts  $e\,{\cal U}_c$, we write the boundary condition in
the form
\be\label{bc}
\phi(0)-\phi(L)-V=[\mu(L)-\mu(0)]/e\equiv{\cal U}_c.
\ee
Equations (\ref{curgen}) and (\ref{bc}) yield the desired current-voltage
characteristics (CVC),
\be\label{CVC}
j L=2\sigma_\eta\, {\cal U}_c+2\sigma V,
\ee
where $2\sigma=2\sigma_\eta+2e^2\nu_0 {\cal D}$ is the total conductivity.
We emphasize that the CVC retains the form (\ref{CVC}) for arbitrary microwave power 
(provided the transport coefficients $\sigma_\eta$ and ${\cal D}$ are calculated 
to all orders in ${\cal P}$). Also, Eq.~(\ref{CVC}) is 
applicable in the case when the microwave-induced redistribution of 
carriers is significant, $\sigma_\eta\sim\sd$, see Eq.~(\ref{etaP}) 
[provided the relative change of the electron density across the sample remains small].
The CVC (\ref{CVC}) is modified only when the linear approximation 
with respect to the dc field breaks down. For such strong dc fields,
 the transport coefficients $\sigma_\eta$ and ${\cal D}$ become field- and coordinate-dependent
(and, therefore, are no longer uniquely defined). Only such strong dc field makes 
important the details of the electrochemical and electrostatic potential 
distribution in the interior of the sample, which necessitates the full solution 
of the Poisson and continuity equations with the boundary conditions (\ref{bc}).

\subsection{Photocurrent and photovoltage}\label{ss53}
\begin{figure}[ht]
\centerline{ 
\includegraphics[width=1.00\columnwidth]{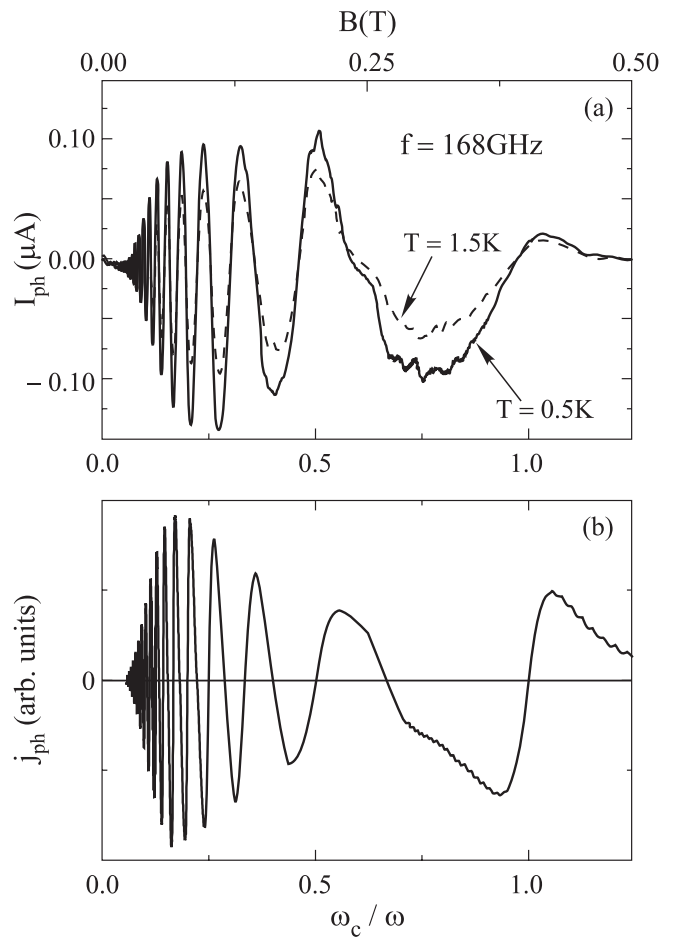}}
\caption{(a) Photocurrent between the Corbino-like internal and strip-like external contact (see Fig.~\ref{fig3}c) vs. magnetic field measured in experimental setup of Ref.~\onlinecite{dorozhkin09} for $T=1.5~{\rm K}$ (solid line) and $T=0.5~{\rm K}$ (dashed line); (b) Photocurrent [calculated using numerical solution of the SCBA equations according to Eqs.~(\ref{f}) and (\ref{jin}), see  Ref.~\onlinecite{crossover} for details] for $\omega\tq=10$ and for a linear polarization of the microwaves. Here we took into account the screening of the incoming radiation by the 2DEG which results in a strong $B$-dependence of the internal microwave field $E_\w$ [entering Eq.~(\ref{Ew})] in the vicinity of the cyclotron resonance.\cite{crossover}}
 \label{fig2}
 \end{figure}

\noindent
If the geometry of two contacts is identical and the difference of the contact potentials is zero, ${\cal U}_c=0$, 
the CVC (\ref{CVC}) reproduces the Ohm law in the bulk, $j=2\sigma V/L$.
Here $\sigma$ contains the displacement and inelastic contributions to the MIRO,
Eqs.~(\ref{sigmaeta}), (\ref{sigmadis}), (\ref{sigmain}), and (\ref{Ddis}), reproducing the
results of previous calculations.\cite{VA04,KV08,DMP03,DVAMP0405,DMP07,DKMPV}
An asymmetric contact configuration results in a non-zero average electric field $\overline{E}= {\cal U}_c/L$
inside the sample in the absence of the bias voltage, $V=0$. 
In the presence of the microwave induced anomalous conductivity, $\sigma_\eta\neq0$, 
the built-in electric field $\overline{E}= {\cal U}_c/L$ leads to the photocurrent at zero bias voltage,
\be\label{jph}
j_{\rm ph}\equiv j|_{V=0}=2\sigma_\eta\, {\cal U}_c/L,
\ee
or, in the open circuit, to the photovoltage
\be\label{Vph}
V_{\rm ph}\equiv V|_{j=0}=-\frac{\sigma_\eta}{\sigma}{\cal U}_c,
\ee
as observed in the experiment.\cite{dorozhkin09} Two experimental traces of the photocurrent for different temperatures 
are shown in Fig.~\ref{fig2}a. Figure \ref{fig2}b illustrates the inelastic contribution (\ref{sigmain}) to the anomalous conductivity (\ref{sigmaeta}), which demonstrates an excellent agreement between the theory and experiment. A
typical small shift of the zeros of the photocurrent from the integer and
half-integer values of $\w/\wc$ in experimental traces is similar to observations~\cite{zudov04,mani04} for MIRO and
can be attributed\cite{zudov04} to a
slight deviation of the electron effective mass from the standard
value $m=0,067m_0$ used in Fig.~\ref{fig2}a.

In the case of overlapping LLs,  the phase and the form of the photocurrent
oscillations is identical for the displacement and inelastic contributions to
$\sigma_\eta$, see Eqs.~(\ref{sigmaeta}),(\ref{sigmadis}), (\ref{sigmain}), and (\ref{RROvLLs}).
Therefore, one can distinguish between them only owing to a strong temperature dependence of 
the inelastic scattering rate. The temperature dependence in Fig.~\ref{fig2}a shows that the inelastic contribution 
to the anomalous conductivity $\sigma_\eta$ is substantial. At the same time, this dependence is weaker than 
$\sigma_\eta^{\rm in}\propto \tin\propto T^{-2}$ predicted by the
theory\cite{DVAMP0405} at the leading order in both the
dc and microwave fields, see Eq.~(\ref{sigmain}). The weaker $T$-dependence can
be attributed either to a strong admixture of the
displacement contribution\cite{KV08,DKMPV,zudov09} at $T=1.5 K$  or to nonlinear
effects\cite{VA04,KV08,DVAMP0405,DMP07} (in the microwave power or
in the dc field). Alternatively, it can be the manifestation of a noticeable
heating of the electron gas\cite{classical,DMP07} at $T=0.5 K$, since
the inelastic scattering time is a function of the electron temperature rather
than the bath (phonon) temperature.

\begin{figure}[ht]
\centerline{ 
\includegraphics[width=0.80\columnwidth]{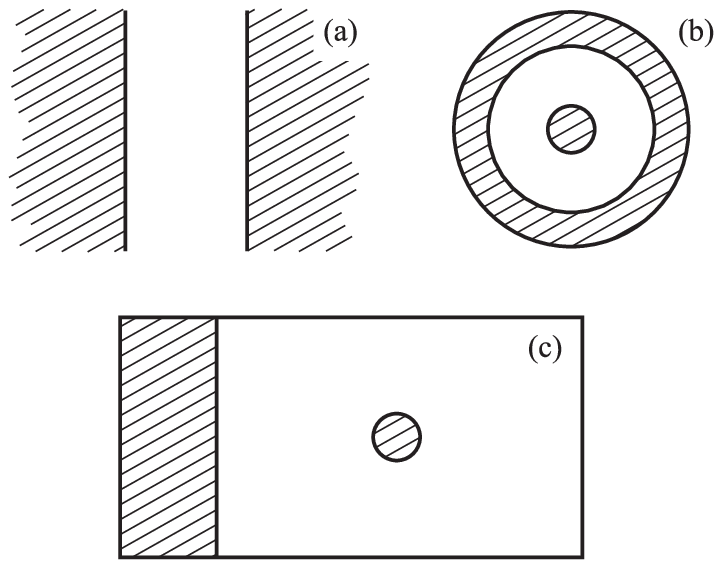}}
\caption{Illustration of different contact geometries. (a) One--dimensional (1D) geometry: 2DEG between two long strip-like contacts. (b) Corbino geometry: as long as the axial symmetry is preserved, can be reduced to the 1D case. (c) Combined geometry 
used in the experiment of Ref.~\onlinecite{dorozhkin09}. The experiment shows that the photogalvanic signal forms near the internal contact having the Corbino geometry while the external strip-like contact plays no role. In such conditions, the theory for geometry (b) is also applicable to the case (c).}
 \label{fig3}
 \end{figure}
Let us emphasize that the analysis of the 1D
geometry considered above (Fig.~\ref{fig3}a) is directly applicable to the case
of the Corbino geometry (Fig.~\ref{fig3}b) for which the Hall conductivity does not enter
the local relation (\ref{genj}). In the case of Corbino geometry, the current density 
$j(r)\propto r^{-1}$ is inversely proportional to the distance from the center, so that the total current $J=2\pi rj(r)$ is conserved. Therefore, integrating both parts of Eq.~(\ref{genj}) along a contour connecting two
contacts at $r=a$ and $r=b$ one obtains a modified CVC in the form
\be\label{CVCcorbino}
J\frac{1}{2\pi}\ln\frac{b}{a}=2\sigma_\eta\, {\cal U}_c+2\sigma V.
\ee
Comparison of Eqs.~(\ref{CVCcorbino}) and (\ref{CVC}) shows that results for 1D geometry (Fig.~\ref{fig2}a)
transform into results for Corbino geometry (Fig.~\ref{fig2}b) after replacement $jL\to J\frac{1}{2\pi}\ln\frac{b}{a}$.
In both cases of 1D and Corbino geometry, the photo-galvanic signals result from non-zero average built-in electric field which requires either 
different work functions of metallic contacts or different electron densities under capacitively coupled
gated probes.\cite{dorozhkin09} A further possible source of the asymmetry is different geometry of the contacts,
e.g. the Corbino-like geometry of the internal contact and the strip-like
geometry of the external contact located on the perimeter of the
sample, see Fig.~\ref{fig2}c (as in the part of the experiment
of Ref.~\onlinecite{dorozhkin09} that utilized heavily doped ohmic
contacts). For such geometry, the photo-galvanic signals were
shown~\cite{dorozhkin09} to be formed in the vicinity the internal
Corbino-like contact and the above consideration should be valid
if one puts the difference of work functions of the doped internal contact and
2DEG instead of ${\cal U}_c$ in Eqs.~(\ref{CVC}), (\ref{jph}), and (\ref{Vph}).

\subsection{Nonlinear effects in the photovoltage; Photoresistance}\label{ss54}\noindent
The magnetooscillations in the photocurrent (\ref{jph}) are fully determined by the anomalous conductivity $\sigma_\eta$ and, therefore, 
the oscillations are symmetric with respect to the average value $j=0$, see Fig.~\ref{fig2}. By contrast, 
the experimental traces of the photovoltage oscillations show\cite{dorozhkin09} a strong asymmetry with respect
to $V_{\rm ph}=0$ value. This asymmetry is due to additional microwave-induced oscillations in the denominator $\sigma=\sd+\sigma_\eta+e^2\nu_0 {\cal D}_{\rm ph}$ of Eq.~(\ref{Vph}). From previous studies of the MIRO\cite{VA04,DVAMP0405}  it is known that contributions to $\sigma_\eta$
of second order in the microwave power are still small when the magnitude of the first-order terms approaches the dark conductivity $\sd$.
This legitimates the use of Eq.~(\ref{Vph}) in the nonlinear regime. Neglecting inessential correction ${\cal D}_{\rm ph}$ [which is a factor $\sim\pi\w/\wc$ smaller than the displacement contribution (\ref{sigmadis}) to $\sigma_\eta$ even in the absence of the inelastic contribution (\ref{sigmain})], one can rewrite Eq.~(\ref{Vph}) as
\be\label{Vph1}
V_{\rm ph}\simeq\frac{-1}{1+\sd/\sigma_\eta}\,{\cal U}_c.
\ee
Equation (\ref{Vph1}) explains a strong asymmetry of the photovoltage oscillations observed in the experiment.\cite{dorozhkin09}
Further, the nonlinearity of Eq.~(\ref{Vph1}) makes possible the experimental determination of the value of the contact potential difference ${\cal U}_c$, since in the formal limit $|\sigma_\eta|\gg\sd$ one has simply $ V_{\rm ph}=-{\cal U}_c$. 

Apart from the photocurrent and photovoltage, one can measure the two-point differential photoresistance $R_{\rm ph}=\partial V/\partial I$ by driving a small current through the sample. Such measurements were also done in Ref.~\onlinecite{dorozhkin09}, and the results were compared to the ratio $V_{\rm ph}/I_{\rm ph}$ taken from two independent measurements of $V_{\rm ph}$ and $I_{\rm ph}$. The comparison demonstrated very good agreement, as can be expected from Eq.~(\ref{CVC}) giving
\be\label{Rph1}
\frac{\partial V}{\partial j}=\frac{L}{2\sigma}
\ee
and Eqs.~(\ref{jph}) and (\ref{Vph}) yielding 
\be\label{Rph2}
-\frac{V_{\rm ph}}{j_{\rm ph}}=\frac{L}{2\sigma}\,.
\ee
The measured photoresistance showed clear magnetooscillations with the phase opposite to the MIRO. The phase shift of oscillations by $\pi$ is in agreement with 
Eqs.~(\ref{Rph1}), (\ref{Rph2}) predicting $R_{\rm ph}\propto\sigma^{-1}$, which should be compared with
$\rho_{\rm xx}\propto\sigma$ in conventional magnetoresistivity measurements of the MIRO.\cite{zudov01,ye01,mani02,zudov03,yang03,dorozhkin03}

\section{Conclusion}\label{s6}\noindent 
Summarizing, we have presented a quantum transport theory for a 2DEG
in high Landau levels illuminated by the microwave radiation in the presence of
a spatially inhomogeneous dc electric field. The theory explains the 
microwave-induced photocurrent and photovoltage oscillations observed in
the recent experiment.\cite{dorozhkin09} 

We have shown that in an irradiated sample the Landau quantization leads to violation of the Einstein relation between the
dc conductivity and the diffusion coefficient. As a result, a non-zero average electric
field leads to the electric current which is
not compensated by the diffusion flow even for the electrochemical potential
remaining constant in space. The experimental observation of the effect requires 
an asymmetry (for instance, in contact geometry, as in Ref.~\onlinecite{dorozhkin09}, 
or in material composition of two contacts) which determines the direction of the current. 
At the same time, the obtained current-voltage characteristics are shown to 
be independent of detailed potential profile in the sample provided 
the relative change in the electron density across the sample remains small.

The effects discussed in this work should also play an
essential role for the transport in the zero resistance
states.\cite{mani02,zudov03,yang03,dorozhkin03} 
In this regime, the uniform charge and
field distributions become electrically unstable.\cite{AAM03} The system breaks into current domains and peculiarities
of the transport properties of the inhomogeneous system become of central importance.

We thank D.N.~Aristov, A.L. Efros, I.V.~Gornyi, M.~Khodas, K.~von~Klit\-zing, J.H.~Smet, M.G.~Vavilov, and
M.A.~Zudov for discussions. 
The experimental results in Fig.~\ref{fig2}a are presented with kind permission of J.H.~Smet and K.~von~Klit\-zing. This work was
supported by INTAS Grant
No.~05-1000008-8044, by the DFG-CFN, by the DFG, and by the RFBR.


\begin{thebibliography}{1}
\bibitem{etanote}{Well-defined quantities in the out-of-equilibrium steady state are the local electron density $n(x)$ and the electrostatic potential $\phi(x)$. In the high-$T$ limit, $2\pi^2 T/\wc\gg 1$, the influence of the Landau quantization on the local chemical potential can be neglected, $\mu(x)=(2\nu_0)^{-1} n(x)$ (for smooth variations of $\mu(x)$ and $n(x)$, see Sec.~\ref{ss43}. Since $\mu/n={\rm const}$, we define the local electrochemical potential as $\eta(x)\equiv\mu(x)+e\phi(x)=(2\nu_0)^{-1} n(x)+e\phi(x)$ and use this definition for a weakly nonequilibrium 2DEG.}
\bibitem{note-mechanisms} As long as corrections to the Hall
conductivity are not considered, contributions to the photocurrent involving higher temporal and angular harmonics of the distribution function can be neglected.\cite{DMP07,DKMPV}
\end{thebibliography}
\end{document}